\documentstyle[preprint,aps]{revtex}

\newcommand{\comm}[1]{\left[#1\right]}
\newcommand{\boldnabla}{\mbox{\boldmath$\nabla$}}

\newcommand{\boldsigma}{\mbox{\boldmath$\sigma$}}
\newcommand{\boldtau}{\mbox{\boldmath$\tau$}}

\begin{document}
\draft

\title{Variational Monte Carlo Calculations of $^3$H and $^4$He with a
Relativistic Hamiltonian - II}

\author{J.L. Forest and V.R. Pandharipande}
\address{Department of Physics, University of Illinois at Urbana-Champaign,
         1110 W. Green St., \\
Urbana, IL 61801}
\author{J. Carlson}
\address{Theoretical Division, Los Alamos National Laboratory, MS B238,
         Los Alamos, NM 87545}
\author{R. Schiavilla}
\address{CEBAF Theory Group, 12000 Jefferson Av., Newport News, VA 23606, \\
and \\
Department of Physics, Old Dominion University, Norfolk, VA 23529}

\date{Sept. 22, 1994}
\maketitle
\begin{abstract}
In relativistic Hamiltonians the two-nucleon interaction is expressed as
a sum of $\tilde{v}_{ij}$, the interaction in the ${\bf P}_{ij}=0$ rest
frame, and the ``boost interaction'' $\delta v({\bf P}_{ij})$ which
depends upon the total momentum ${\bf P}_{ij}$ and vanishes in the
rest frame. The $\delta v$ can be regarded as a sum of four terms:
$\delta v_{RE}$, $\delta v_{LC}$, $\delta v_{TP}$ and $\delta v_{QM}$;
the first three originate from the relativistic energy-momentum relation,
Lorentz contraction and Thomas precession, while the last is purely
quantum. The contributions of $\delta v_{RE}$ and $\delta v_{LC}$
have been previously calculated with the variational Monte Carlo method
for $^3$H and $^4$He. In this brief note we report the results of
similar calculations for the contributions of $\delta v_{TP}$ and
$\delta v_{QM}$. These are found to be rather small.
\end{abstract}
\newpage

Recently we reported\cite{carlson} results of variational Monte Carlo
calculations of $^3$H and $^4$He with a relativistic Hamiltonian based
on the work of Foldy\cite{foldy}, Krajcik and Foldy\cite{krajcik} and
Friar\cite{friar}. This Hamiltonian has the form:
\begin{equation}
H=\sum_i\left[\left(m^2+p_i^2\right)^{1/2}-m\right]+\sum_{i<j}\left[\tilde{v}_{ij}+\delta v({\bf P}_{ij})\right]+\sum_{i<j<k}V_{ijk},
\end{equation}
where ${\bf p}_i$ label momenta of particles, and
${\bf P}_{ij}={\bf p}_i+{\bf p}_j$ is the total momentum of the pair $ij$.
The two-nucleon interaction $\tilde{v}_{ij}$ is obtained by fitting the
scattering data in the ${\bf P}_{ij}=0$ frame. The boost interaction
$\delta v({\bf P}_{ij})$ is zero when ${\bf P}_{ij}=0$, and is generally
given by:
\begin{equation}
\delta v({\bf
P}_{ij})=-\frac{P_{ij}^2}{8m^2}\tilde{v}_{ij}+\frac{1}{8m^2}\comm{{\bf
P}_{ij}\cdot{\bf r}_{ij}{\bf
P}_{ij}\cdot\boldnabla_{ij},\tilde{v}_{ij}}+\frac{1}{8m^2}\comm{\left(\boldsigma_i-\boldsigma_j\right)\times{\bf P}_{ij}\cdot\boldnabla_{ij},\tilde{v}_{ij}}
\end{equation}
up to order $P_{ij}^2/m^2$. Only the first two terms of this
$\delta v({\bf P}_{ij})$ were considered in ref. [1]. The last term, having
$(\boldsigma_i-\boldsigma_j)$, does not have diagonal matrix elements
in eigenstates of $S^2=(\boldsigma_i+\boldsigma_j)^2$. Hence it was
neglected in [1]. The Urbana model VII of $V_{ijk}$ is used, and its
boost correction $\delta V_{ijk}({\bf P}_{ijk})$ is neglected. This correction
is zero for $^3$H in its rest frame, and in $^4$He it is expected to
contribute much less than the $\delta v({\bf P}_{ij})$.

In the present work we calculate the expectation value of the
$(\boldsigma_i-\boldsigma_j)$ term in $\delta v({\bf P}_{ij})$. This term
can couple the dominant two-nucleon $T, S=1,0$ and $0,1$ waves in the
wave function of $^3$H and $^4$He to the small P-waves having $T,S=1,1$
and $0,0$ respectively. The $\tilde{v}_{ij}$ has fourteen terms like those
of the Urbana $v_{14}$ interaction\cite{lagaris}. The first six of these
have operators
$(1,\boldsigma_i\cdot\boldsigma_j,S_{ij})\otimes(1,\boldtau_i\cdot\boldtau_j)$,
and are denoted by $\tilde{v}_{6,ij}$:
\begin{eqnarray}
\tilde{v}_{6,ij} & = &
v_c(r_{ij})+v_\sigma(r_{ij})\boldsigma_i\cdot\boldsigma_j+v_t(r_{ij})S_{ij}
\nonumber \\
& &
+\left[v_\tau(r_{ij})+v_{\sigma\tau}(r_{ij})\boldsigma_i\cdot\boldsigma_j+v_{t\tau}(r_{ij})S_{ij}\right]\boldtau_i\cdot\boldtau_j.
\end{eqnarray}
The $\tilde{v}_{6,ij}$ gives $>98\%$ of the $\langle\tilde{v}_{ij}\rangle$
in $^3$H and $^4$He, therefore we approximate the $\tilde{v}_{ij}$
in the $(\boldsigma_i-\boldsigma_j)$ term of $\delta v({\bf P}_{ij})$ by
$\tilde{v}_{6,ij}$.

The commutator can be written as:
\begin{equation}
\frac{1}{8m^2}\comm{\left (\boldsigma_i-\boldsigma_j\right )\times{\bf
P}_{ij}\cdot\boldnabla_{ij},\tilde{v}_{6,ij}}=\delta v_{TP}({\bf
P}_{ij})+\delta v_{QM}({\bf P}_{ij}),
\end{equation}
where
\begin{equation}
\delta v_{TP}({\bf
P}_{ij})=\frac{1}{8m^2}\left(\boldsigma_i-\boldsigma_j\right)\times{\bf
P}_{ij}\cdot\left(\boldnabla_{ij}\tilde{v}_{6,ij}\right),
\end{equation}
and $\delta v_{QM}({\bf P}_{ij})$ contains terms that come from the commutator
of $(\boldsigma_i-\boldsigma_j)$ with the spin operators in $\tilde{v}_{6,ij}$.
The $\delta v_{TP}({\bf P}_{ij})$ originates from the classical Thomas
precession\cite{jackson,forest}. The precession of the spin ${\bf s}_i$ in
the frame moving with velocity ${\bf P}_{ij}/2m$ is given by
$-\boldnabla_{ij}\tilde{v}_{ij}\times{\bf P}_{ij}/4m^2$ up to order
$1/m^2$. Thus the Thomas precession potential for particle $i$ is:
\begin{equation}
-\frac{1}{2}\boldsigma_i\cdot\frac{\boldnabla\tilde{v}_{ij}\times{\bf
P}_{ij}}{4m^2}=\frac{1}{8m^2}\boldsigma_i\times{\bf P}_{ij}\cdot\left
(\boldnabla_{ij}\tilde{v}_{ij}\right ).
\end{equation}
Both particles have same velocity due to their center of mass motion, but
their accelerations due to $\tilde{v}_{ij}$ are equal and opposite. Therefore
the Thomas precession potential for the particle $j$ is
$-\boldsigma_j\times{\bf P}_{ij}\cdot\left (\boldnabla_{ij}\tilde{v}_{ij}\right
)/8m^2$, and together with (6) it makes up the $\delta v_{TP}({\bf P}_{ij})$.
After some algebra we obtain:
\begin{eqnarray}
\delta v_{TP}({\bf P}_{ij}) & = & \frac{1}{8m^2r}\left[\left
(v_c^{\prime}-v_{\sigma}^{\prime}+v_t^{\prime}+3\frac{v_t}{r}\right){\bf
P}\cdot{\bf r}\times(\boldsigma_i-\boldsigma_j)\right. \nonumber \\
& & \left.
-i\left(2v_{\sigma}^{\prime}+v_t^{\prime}+3\frac{v_t}{r}\right)\left({\bf
P}\cdot\boldsigma_i\ {\bf r}\cdot\boldsigma_j-{\bf P}\cdot\boldsigma_j\ {\bf
r}\cdot\boldsigma_i\right)\right]+\boldtau_i\cdot\boldtau_j\ \mbox{term},
\end{eqnarray}
where $v_x^{\prime}$ denotes $\partial v_x/\partial r$, the $ij$ subscripts
of ${\bf r}$, ${\bf P}$ and $v_x$ are omitted for brevity, and the
$\boldtau_i\cdot\boldtau_j$ term has $v_{\tau}$, $v_{\sigma\tau}$ and
$v_{t\tau}$ in place of $v_c$, $v_{\sigma}$ and $v_t$.

The $\delta v_{QM}({\bf P}_{ij})$ does not have a classical analogue; it is
found to be:
\begin{eqnarray}
\delta v_{QM}({\bf P}_{ij}) & = & \frac{i}{2m^2}(v_t-v_{\sigma})\left({\bf
P}\cdot\boldsigma_i\ \boldsigma_j\cdot\boldnabla-{\bf P}\cdot\boldsigma_j\
\boldsigma_i\cdot\boldnabla\right) \nonumber \\
& & -\frac{3i}{4m^2}\ \frac{v_t}{r^2}\ {\bf P}\cdot{\bf
r}\left(\boldsigma_i\cdot{\bf r}\
\boldsigma_j\cdot\boldnabla-\boldsigma_j\cdot{\bf r}\
\boldsigma_i\cdot\boldnabla\right) \nonumber \\
& & -\frac{3i}{4m^2}\ \frac{v_t}{r^2}\left({\bf P}\cdot\boldsigma_i\ {\bf
r}\cdot\boldsigma_j-{\bf P}\cdot\boldsigma_j\ {\bf
r}\cdot\boldsigma_i\right){\bf r}\cdot\boldnabla \nonumber \\
& & +\boldtau_i\cdot\boldtau_j\ \mbox{terms}
\end{eqnarray}
from eq. (4).

It is convenient\cite{forest} to express $\delta v({\bf P}_{ij})$ given
by eq. (2) as:
\begin{equation}
\delta v({\bf P}_{ij})=\delta v_{RE}({\bf P}_{ij})+\delta v_{LC}({\bf
P}_{ij})+\delta v_{TP}({\bf P}_{ij})+\delta v_{QM}({\bf P}_{ij}).
\end{equation}
Its first term:
\begin{equation}
\delta v_{RE}({\bf P}_{ij})=-\frac{P_{ij}^2\ \tilde{v}_{ij}}{8m^2}
\end{equation}
comes from the relativistic energy, and the second:
\begin{equation}
\delta v_{LC}({\bf P}_{ij})=\frac{1}{8m^2}{\bf P}_{ij}\cdot{\bf r}_{ij}\ {\bf
P}_{ij}\cdot\left(\boldnabla_{ij}\,\tilde{v}_{ij}\right)
\end{equation}
from Lorentz contraction. The $\comm{{\bf P}_{ij}\cdot{\bf r}_{ij}{\bf
P}_{ij}\cdot\boldnabla_{ij},\tilde{v}_{ij}}$ can have terms in addition to
those in
$\delta v_{LC}$ when $\tilde{v}_{ij}$ depends upon the relative momentum
${\bf p}_{ij}$. These terms are to be regarded as a part of $\delta v_{QM}$.
However, they vanish when $\tilde{v}_{ij}$ is approximated with
$\tilde{v}_{6,ij}$.

The expectation values of $\delta v_{TP}({\bf P}_{ij})$ and
$\delta v_{QM}({\bf P}_{ij})$ are calculated with the variational
wave function of ref. [1] using the Monte Carlo methods described
in [1]. The results are tabulated in table I along with others of
interest from [1]. The contributions of $\delta v_{TP}$ and
$\delta v_{QM}$ are much smaller than those of $\delta v_{RE}$
and $\delta v_{LC}$ as expected. These contributions would be exactly
zero if there were no two-nucleon P-waves in these nuclei.

Stadler and Gross\cite{stadler} have also estimated these contributions
in $^3$H with a different method and obtained similar results.

The authors would like to thank Dr. J. L. Friar for illuminating
discussions. The work of JLF and VRP is partly supported by the
U.S. National Science Foundation via grant PHY--89--21025, that of
JC and RS was performed under the auspices of the U.S. Department
of Energy.

\begin{table}
\caption{Expectation values in MeV}
\begin{tabular}{ddd}
&$^3$H&$^4$He\\
\tableline
$\langle\displaystyle{\sum_i}(m^2+p_i^2)^{1/2}-m\rangle$&  48.7(2)  &  105.0(6)
  \\
$\langle\displaystyle{\sum_{i<j}}\tilde{v}_{ij}\rangle$& -55.9(2)  & -127.4(5)
 \\
$\langle\displaystyle{\sum_{i<j<k}}\tilde{v}_{ijk}\rangle$& -1.21(2)  &
-5.43(15)   \\
$\langle\displaystyle{\sum_{i<j}}\delta v_{RE}({\bf P}_{ij})\rangle$&  0.23(2)
&  1.17(3)    \\
$\langle\displaystyle{\sum_{i<j}}\delta v_{LC}({\bf P}_{ij})\rangle$&  0.10(1)
&  0.53(1)    \\
$\langle\displaystyle{\sum_{i<j}}\delta v_{TP}({\bf P}_{ij})\rangle$&  0.016(2)
&  0.074(4)   \\
$\langle\displaystyle{\sum_{i<j}}\delta v_{QM}({\bf P}_{ij})\rangle$& -0.004(2)
&  -0.014(4)  \\
$\langle H\rangle$& -8.07(3)  & -25.90(8) \\
\end{tabular}
\end{table}

\end{document}